**Scalable electrodeposition of liquid metal from an acetonitrile-based electrolyte for highly-integrated stretchable electronics**

Wouter Monnens[†], Bokai Zhang[†], Zhenyu Zhou[†], Laurens Snels[†], Koen Binnemans[‡], Francisco Molina-Lopez[†]*, Jan Fransaer[†]*

[†]KU Leuven, Department of Materials Engineering, Kasteelpark Arenberg 44, P. O. box 2450, B-3001 Leuven, Belgium

[‡] KU Leuven, Department of Chemistry, Celestijnenlaan 200F, P.O. box 2404, B-3001 Leuven, Belgium.

*Corresponding authors:

Email: francisco.molinalopez@kuleuven.be and Jan.fransaer@kuleuven.be

**Abstract**

For the advancement of highly-integrated stretchable electronics, the development of scalable (sub)-micrometer conductor patterning is required. Eutectic gallium indium (EGaIn) is an attractive conductor for stretchable electronics, as its liquid metallic character grants it high electrical conductivity upon deformation. However, its high surface energy precludes patterning it with sub-micron resolution. Herein, we overcome this limitation by reporting for the first time the electrodeposition of EGaIn. We use a non-aqueous acetonitrile-based electrolyte that exhibits high electrochemical stability and chemical orthogonality. The electrodeposited material led to low-resistance lines that remained stable upon (repeated) stretching to a100% strain. Because electrodeposition benefits from the resolution of mature nanofabrication methods used to pattern the base metal, the proposed "bottom-up" approach achieved a record-high density integration of EGaIn regular lines of 300 nm half-pitch on an elastomer substrate by plating on a gold seed layer pre-patterned by nanoimprinting. Moreover, vertical integration was enabled by filling high aspect ratio vias. This capability was conceptualized by the fabrication of an omnidirectionally stretchable 3D electronic circuit, and demonstrates a soft-electronic analogue of the stablished damascene process used to fabricate microchip interconnects. Overall, this work proposes a simple route to address the challenge of metallization in highly-integrated (3D) stretchable electronics.

**Introduction**

Stretchable electronics is an emerging and extensively researched field that encompasses the development of soft elastic electronics devices which can be interfaced with biological tissue.[1,2] These devices can be engineered to conform to human skin, enabling them to be utilized as wearable or implantable sensors and as apparatuses for health monitoring and (bio)data processing.[3,4] Several methods have been developed to fabricate stretchable devices: lamination of flexible thin-film materials on pre-strained substrates, complex patterning of thin films as meanders, synthesis of intrinsically stretchable materials, or a combination of all.[5] In applications involving prolonged periods of induced strain, devices made entirely of stretchable materials are more durable than their counterparts made of rigid materials, which are susceptible to concentrated stress concentrations at the soft-rigid interfaces. Liquid metals such as eutectic alloys of In, Ga and Sn, have been extensively used as electrodes and interconnects in stretchable electronics due to their high conductivity compared with stretchable conducting polymers ($10^6$ vs. $10^5$ S m$^{-1}$) and their ability to withstand strains up to 700%.[6,7]

The patterning of intrinsically stretchable materials is usually done from solution using low resolution methods such as inkjet printing, stamping, or spray coating.[8] There is currently a lack of patterning methods that enable the high component/interconnect density with nanoscale resolution typical of rigid electronics, precluding the development of high-performing, complex stretchable circuits. Recently, Y-Q Zheng *et al.* exploited the patterning strength of photolithography for stretchable electronics by *monolithic optical microlithography*, an ingenious strategy that consists of the chemical modification of stretchable electronic polymers to induce light–triggered solubility modulation, allowing high-density elastic circuits patterned with 2 µm resolution.[9] However, this technique requires modifying the chemistry of the starting polymers, and does not foresee the fabrication of vias to connect different metallization layers, which albeit challenging, is the next logical processing step. Moreover, *monolithic optical microlithography* is not directly applicable to liquid metals, which are one order of magnitude more electrically conductive than conducting polymers. The miniaturization of liquid metals is especially challenging owing to their high surface tension.[10] Attempts to use photolithography-inspired techniques to pattern liquid metal yielded 20 µm resolution,[11,12] whereas direct nozzle printing and hybrid processes combining lithography, stamping and soft nanoimprinting steps have demonstrated patterning with a resolution of a few µm.[13–16] However, sub-micrometer patterning of liquid metal remains elusive. To the best of our knowledge, only one example has been reported by M. Kim *et al*. demonstrating patterning as small as 180 nm via a multi-step non-standard hybrid method involving electron-beam lithography and soft lithography.[17] Although yielding impressive results, the previous work involved a rather complex process flow and was limited to single-line patterning.

To the best of our knowledge, electrodeposition of eutectic gallium indium (EGaIn), or any other binary gallium-indium alloys has never been reported. For the selection of a suitable electrolyte that enables electrodeposition of EGaIn, it is essential to consider the (electro)chemical behavior of gallium and indium in various media. Due to their negative standard reduction potentials ($E^0_{In(III)/In}$ = −0.34 V vs. SHE and $E^0_{Ga(III)/Ga}$ = −0.53 V vs. SHE), the electrodeposition of both metals from aqueous electrolytes can compete with the hydrogen evolution reaction (HER).[18] Especially in acidic aqueous electrolytes, the efficiency of gallium electrodeposition is very low due to concurrent HER, whereas in alkaline electrolytes, its electrochemical behavior is complex, involving unstable species and the formation of gallium(III) oxide.[19–21] Non-aqueous electrolytes can exhibit higher electrochemical stabilities than water, yielding higher electrodeposition efficiencies for gallium. However, these electrolytes can support the formation of intermediate monovalent species, both for indium and gallium, which are typically unstable and undergo a disproportionation reaction.[22–27] Ideally, the formation of these monovalent species should be avoided as it complicates the overall process.

In this work, we report on the electrodeposition of EGaIn liquid metal from an acetonitrile-based electrolyte for application in high-density integration of stretchable electronics. The use of an acetonitrile-based bath allowed us to address all the typical bottlenecks associated to the electrodeposition of binary gallium-indium alloys. Acetonitrile exhibits a broad electrochemical window and is especially resilient against reduction. Furthermore, it allows dissolution of high concentrations (> 0.5 M) of $InCl_3$ and $GaCl_3$, resulting in high cathodic currents and therefore high deposition rates. It is shown that the formation of monovalent species is also very low by means of rotating ring disk electrode measurements. Last yet importantly, being a polar solvent, acetonitrile does not attack typical elastomers like PDMS or styrene ethylene butylene styrene (SEBS), and is compatible with stretchable electronic polymers, which are based on conjugated molecules soluble usually in non-polar solvents.[28]

Due to the fluidic and metallic nature of EGaIn, our process rendered robust stretchable liquid metal lines with dimensions varying from millimeter to sub-micron scale displaying a conductivity of $2 \times 10^6$ S m$^{-1}$ that was maintained up to a strain of 100% (strain limited by the substrate fracture rather than by the brake-up of the liquid metal lines) and showed stability after 300 (50%) strain cycles. The proposed technique presents a breakthrough in the integration of stretchable electronics because its "bottom-up" nature bypasses the challenges of fine patterning associated to the high surface tension of liquid metal. Indeed, our process shifts the patterning complexity from the liquid metal to the seed layer, which can be deposited using traditional nanofabrication techniques such as stencil/nanoimprint/photo-lithography and evaporation/sputtering. This allows scalability, the definition of complex patterns, and the miniaturization of features well beyond the state of art, even on elastomeric substrates. We showcase this point by a simple and low-cost process in which the nanopatterning of a thin film of sputtered gold is achieved by nanoimprinting a polydimethylsiloxane (PDMS) substrate and lifting-off using adhesive tape. After electroplating, the gold seed layer yields EGaIn lines on PDMS with an unprecedented resolution of 300 nm (300 nm width and 300 nm spacing), paving the way to high-density stretchable circuitry. Furthermore, electrodeposition enables vertical 3D integration by filling high aspect ratio features in a void-free manner, which can be challenging for the "top-down" techniques typically used for depositing liquid metals. We drew inspiration from the damascene copper electroplating process used by the semiconductor industry to fabricate interconnects for the back end of line (BEOL) of integrated circuits, and demonstrate the filling of through-elastomer stretchable vias with 8:1 aspect ratio that served to power integrated microLEDs under mechanical strain.

**Results and discussion**

*Substrate preparation and electrochemistry*

A simple method was developed to fabricate stretchable substrates on which liquid metal lines could be grown by electrodeposition. The process flow is depicted in Figure 1a. Thin PDMS pieces supported on glass substrates were covered by a plastic foil shadow mask with a cut-out (1) and subsequently sputtered with a thin (~5 nm) gold layer (2). The mask was removed, yielding the desired gold seed layer on PDMS (3). A droplet of silver paste ink (~1 ml) was then added to one end of the seed layer and dried at 40°C in a vacuum chamber for 24 hours (4). This droplet served as an electrical connection point during the electrodeposition process. Electrodeposition of EGaIn was subsequently carried out in a three-electrode setup as shown in Figure 1b. A cyclic voltammogram (CV) was first recorded for acetonitrile with 0.1 M of tetrabutylammonium chloride ([TBA][Cl]) as background salt (Figure 1c, black line), to determine the potential window of operation in which electrodeposition of EGaIn can be carried out. This chloride-based background salt was chosen to reveal possible electrochemical reactions that could be attributed to the chloride anions present in the gallium and indium precursors, which are metal(III) chlorides. The cathodic limit and anodic limit are positioned at −2.75 V vs. $Ag^+/Ag$ and +0.97 V vs. $Ag^+/Ag$, respectively, indicating a wide electrochemical window of 3.72 V. The observed cathodic limit of acetonitrile (which is relevant for the deposition efficiency) occurs at large overpotentials analogous to those described in the literature (using the same reference electrode), and it is only the anodic limit which is shifted to lower anodic overpotentials. This shift is due to the oxidation of chloride (originating from added TBACl) to chlorine, narrowing the window of electrolyte. While the high stability towards cathodic breakdown is beneficial for electrodeposition of gallium and indium in a wide cathodic overpotential range, the limited anodic stability is not of concern, as we are not interested in stripping the deposited liquid metal.

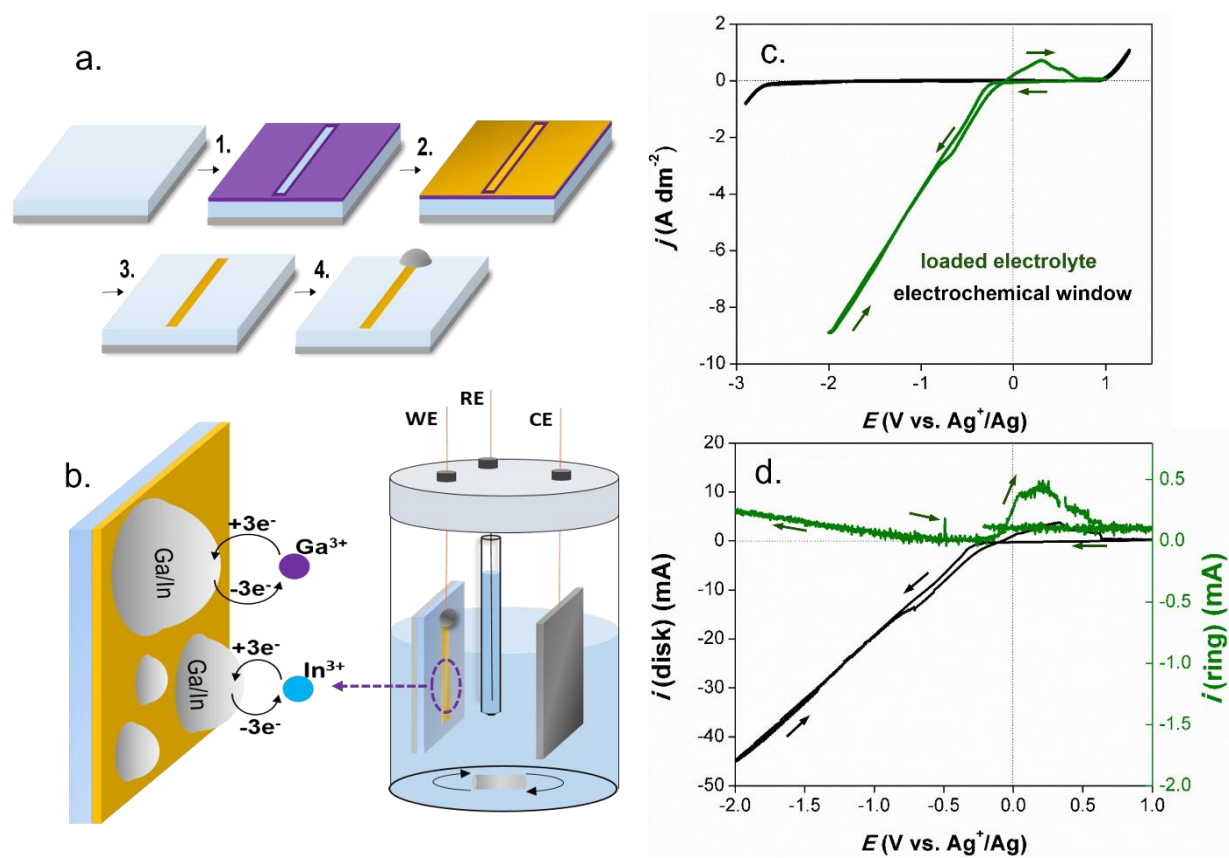

Figure 1: a) Sketch of the stepwise fabrication of macroscopic stretchable liquid metal patterns. b) Sketch of the applied three-electrode setup and a detailed description of the reactions occurring on the substrate (WE: working electrode; RE: reference electrode; CE: counter electrode). c) CVs of acetonitrile with 0.1 M of TBACl (black line) and acetonitrile with 500 mM of $GaCl_3$ and 87.5 mM of $InCl_3$ (green line), recorded on the PDMS/Au electrodes at room

temperature with a scan rate of 10 mV s$^{-1}$. d) Rotating ring disk electrode (RRDE) measurement where a CV recorded on a gold disk electrode (∅ = 6.5 mm) for acetonitrile with 500 mM of GaCl$_3$ and 87.5 mM of InCl$_3$ at a scan rate of 10 mV s$^{-1}$ at room temperature (black line, left axis) with current response on platinum ring on which +0.3 V vs. Ag$^+$/Ag was applied (green line, right axis). The RRDE was rotated at 400 rpm.

After assessing the suitability of acetonitrile as solvent for gallium and indium electrodeposition, an electrolyte composed of acetonitrile, 500 mM GaCl$_3$, and 87.5 mM InCl$_3$ was prepared. These concentrations corresponded to a gallium/indium precursor ratio of 10/1.75, which allowed to reach the eutectic point in the final electroplated films (75.5 w% gallium and 24.5 w% indium) leading to liquid metal. The recorded CV for this electrolyte is depicted in green in Figure 1C. In the forward scan, an onset in reduction current density takes place at −0.25 V vs. Ag$^+$/Ag. The current density steadily increases, reaching nearly −9 A dm$^{-2}$ at −2.0 V vs. Ag$^+$/Ag. The corresponding processes are ascribed to the reduction of indium(III) to indium and gallium(III) to gallium, respectively. A high concentration of precursors was used to avoid a diffusion-limited reduction since this typically enables a stable deposition process. Moreover, the high current densities lead to high deposition rates of liquid metal coatings. The absence of a limiting current in the CV confirms that the process is not diffusion-controlled. In the reverse scan, the current passes through zero at a −0.07 V vs. Ag$^+$/Ag and makes a small nucleation loop. This feature is indicative of metal deposition. At −0.30 V vs. Ag$^+$/Ag, an oxidation peak occurs, which can be ascribed to the stripping of indium, gallium, or formed indium-gallium alloys. However, the large discrepancy in charge consumed during reduction and stripping implies that not all deposited metal is stripped (charge vs. potential plot is shown in ESI Figure 1s). This was also evident from the metallic shiny grey coating that covered the gold layer after the recording of the CV. Another reason that can contribute to this discrepancy is that while during electrodeposition, three electrons are required per indium(III) or gallium(III) species, stripping could possibly only involve one electron, e.g., in case of stripping of indium and/or gallium metal to monovalent indium(I) and/or gallium(I) species. To determine whether undesired monovalent indium(I) and/or gallium(I) are formed during reduction and stripping, a rotating ring disk electrode (RRDE) measurement was performed.[29,30] In this measurement, a CV was recorded on the central disk electrode, and an anodic overpotential of +0.3 vs. $Ag^+/Ag$ was applied on the outer ring electrode, while the RRDE was rotated at 400 rpm. In principle, the monovalent species formed on the disk during the recording of the CV are spun outwards the ring and detected through oxidation to the trivalent species by the applied anodic overpotential. The corresponding measurement is shown in Figure 1d. The ring current starts to deviate from zero at −1.0 vs. $Ag^+/Ag$, indicating the onset of the formation of monovalent species by two-electron reduction of trivalent species as:

$$\text{DISK reaction: M(III)} + 2e^- \rightarrow \text{M(I) with (M = Ga or In)}$$
$$\text{RING reaction: M(I)} \rightarrow \text{M(III)} + 2e^- \text{ with (M = Ga or In)}$$

(1)

The ring current consistently increases till −0.245 mA at −2.0 vs. $Ag^+/Ag$, whereas the disk current reaches a much larger current of −46 mA. Taking into account the collection efficiency of the RRDE (38 %), and the exchanged electrons (2 electrons on both the disk (M(III) to M(I)) and the ring (M(I) to M(III))), it can be calculated that the disk current corresponding to the formation of the monovalent species only accounts for 1.4 % of the total disk current. Hence, at −2.0 vs. $Ag^+/Ag$, 98.6 % of the total reduction current can be attributed to the electrodeposition of gallium and indium, confirming the suitability of the selected electrolyte system for high-efficiency deposition. In the reverse scan, overlapping with the stripping peak on the disk, a current peak on the ring can be observed. This indicates that stripping of deposited metal involves the formation of monovalent species. The disk peak current reaches 3.76 mA whereas the ring peak current reaches −0.44 mA. Using the same calculation as mentioned above, roughly 31 % of the total stripping current of the disk involves oxidation of the metal to monovalent species. Possibly, during stripping, a part of the formed monovalent species is further oxidized to trivalent species. However, as this work focuses on the electrodeposition, the stripping/oxidation process was not investigated further.

*Conditioning and characterization of electrodeposited EGaIn lines*

Analysis of the CV already indicated that efficient electrodeposition of indium-gallium is feasible. Furthermore, the composition of the gallium-indium alloy could be precisely controlled by the applied cathodic overpotential as shown in Figure 2a. Electrodeposition was carried out in the range of −0.75 V to −2.5 V vs. $Ag^+/Ag$ on gold substrates for 5 minutes using the electrolyte composed of acetonitrile with 500 mM of $GaCl_3$ and 87.5 mM of $InCl_3$. The gallium/indium ratio was quantified using SEM-EDX. A clear trend is observed in which the gallium/indium ratio steadily increases with increasing overpotential. At −2.0 V vs. $Ag^+/Ag$, the composition of EGaIn leading to liquid metal at room temperature is reached. Figure 2b depicts SEM images of the EGaIn deposit. The morphology consists of micro-sized spherical droplets with a broad range of diameters. These droplets, composed of a liquid EGaIn core with a thin gallium oxide shell, are typically found in EGaIn due to its extremely high surface tension ($\approx$ 625 mN $m^{-1}$) of EGaIn.[31] The gallium oxide shell enveloping the spheres spontaneously forms when EGaIn is exposed to $O_2$-containing environments even at concentrations as low as 1 ppm.[32] The thickness of the shell is only 0.5–5 nm, yet it provides stability to the inner liquid content.[33] This (amorphous) gallium oxide shell does not block the further electrodeposition as it permits the reduction of indium and gallium ions on its surface, forming nuclei and ultimately new spheres. This is evident from the small spheres observed on the surfaces of the bigger ones below. This deposition and growth mechanism is assumed to be analogous to that of pure gallium, which was investigated and discussed in a previous work.[24] This spherical morphology, however, is undesirable for use as electrical conductors in circuitry, as the oxide shells increase the electrical resistance. Furthermore, upon stretching the substrate, the neighboring spheres might be pulled from each other, creating voids, and disrupting the conducting path. To tackle this, the oxide shells of the spheres need to be ruptured to make continues liquid metal structures. Various studies investigated the sintering of EGaIn (nano)droplets and reported multiple ways to rupture the oxide shells. These are mechanical sintering, acoustic sintering, laser-induced sintering and thermal sintering.[12,34–37] We took advantage of the compatibility of our process to deposit directly on a stretchable substrate to develop a simple mechanical sintering consisting of stretching and bending the sample a few times. Figures 2b2&3 confirm this synergistic effect of substrate and EGaIn with SEM images showing a partly sintered EGaIn, in which the sample was once bent manually right before imaging leading to the coalescence of most droplets; and a fully sintered sample achieved by few bending repetition, displaying the smooth surface desired for a reliable conductor. It is worth noting that the thin gold seed layer would crack upon stretching, losing its conductivity as a result. However, this is not a problem because after the liquid metal electrodeposition is completed, the gold seed layer is no longer needed. At this point, the liquid metal spreads uniformly above the cracked gold becoming the main contributor to electrical conduction. A picture of the sintered sample alongside a top-view SEM image and elemental map are shown in Figure 2c. The image and map reveal a smooth liquid metal line, consisting of equally distributed gallium and indium. The conductivity of the deposited line was determined to be $2.5 \times 10^6$ S m$^{-1}$, which is in the same order of magnitude as is defined in literature.[38] Both the morphology and the conductivity remained unchanged after repeating stretching and bending.

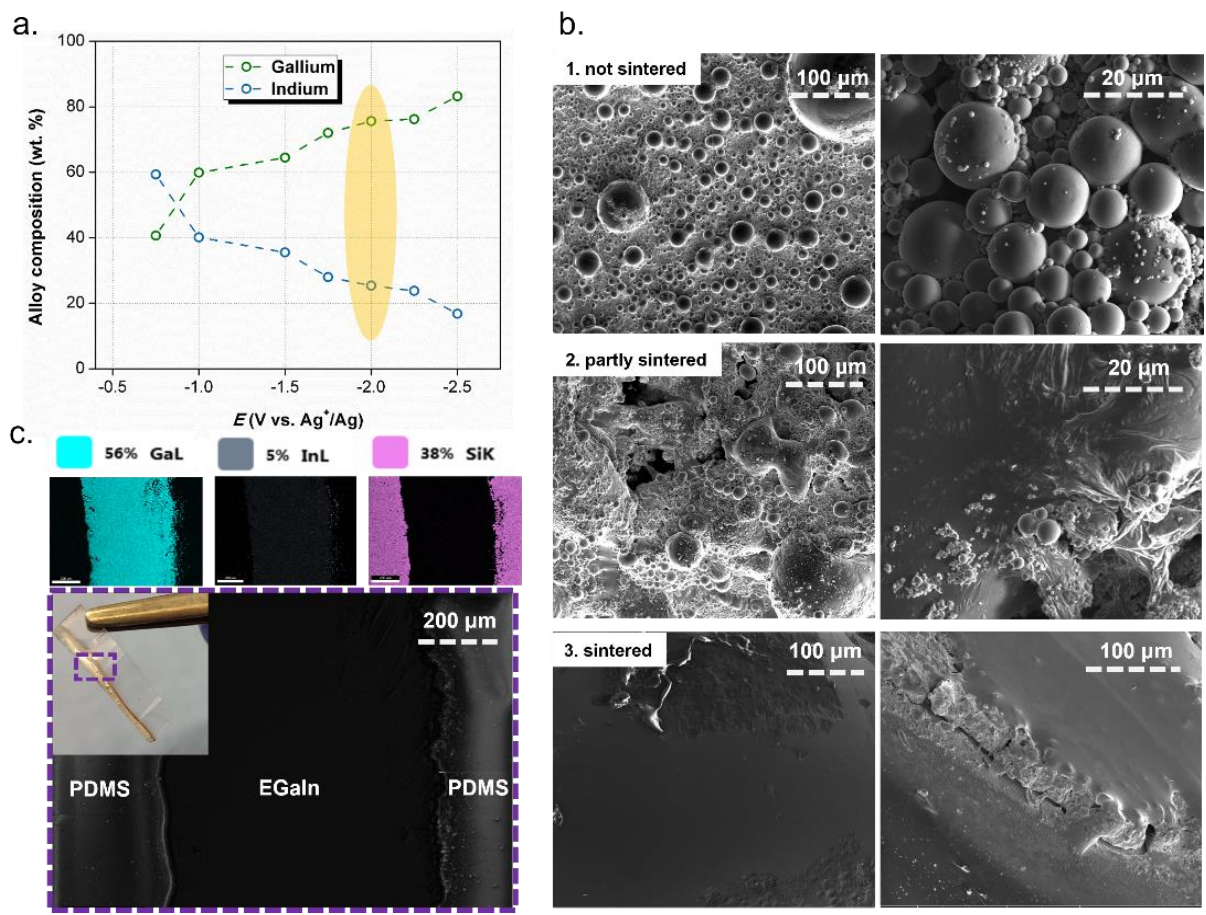

Figure 2: a) InGa alloy composition vs. cathodic overpotential diagram. b) SEM images of (1) not sintered, (2) partly mechanically sintered and (3) fully mechanically sintered EGaIn electroplated by applying −2.00 V vs. Ag$^+$/Ag for 2 min. using an electrolyte composed acetonitrile, 500 mM of GaCl$_3$ and 87.5 mM of InCl$_3$. c) Picture of mechanically sintered sample with SEM image and elemental map of the deposited EGaIn line (the Si corresponds to the uncovered PDMS substrate).

To test the mechanical performance of the deposited EGaIn lines, dogbone-shaped lines plated on a PDMS substrate (Figure 3a) were mounted in a custom-made stretching setup (Figure 3b), and subjected to uniaxial tensile strain up to 100% strain, while measuring their electrical resistance. The relative resistance evolution ($R/R_0$) is plotted in Figure 3c, where $R_0$ (= 0.11 Ω) and $R$ are the resistance in absence and presence of strain, respectively. The curve shows the high stretchability of the electroplated EGaIn lines, which remain highly conductive up to the PDMS substrate fracture at 100 % strain. It can be reasonably suspected that the thin and brittle gold layer used as seed for electroplating, cracks and becomes non-conductive during these strain experiments. Hence, nearly all the electrical conduction comes from the continuous EGaIn film that bridges the crack gold. The EGaIn lines exhibited changes in their geometrical shape upon stretching: as the length increased, the width decreased according to the Poisson's ratio of PDMS (assumed = 0.5). In this case, the PDMS substrate imposes the line width change but the height of the liquid metal film remains free. However, since liquid metal is an incompressible fluid, its volume must remain constant during stretching, which constraints the film thickness to decrease also according to a Poisson's ratio of 0.5 (see eq. S4 and discussion). Then, according to Pouillet's law that dictates that the resistance of a wire is proportional to its length and resistivity, and inversely proportional to its cross-sectional area, the theoretical correlation between $R/R_0$ and strain $\varepsilon$ for an incompressible material with a constant resistivity should be $R/R0 = (1+\varepsilon)^2$ (see eq. S8 and discussion). The measured change in resistance of our liquid metal lines was lower than the theoretical prediction. This behavior has been reported before and can been ascribed to different reasons, being contact resistance a main one for samples with very low value of resistance (usually the case for the highly conducting EGaIn).[39] We estimated the effect of high contact resistance in the value of $R/R_0$ Vs strain (Figure S4), and concluded that to account for the low values registered in the first stretching cycle observed in Figure 3c, the contact resistance should contribute to 90% of the total resistance (i.e. $R_c$ should be ~ 9 times larger than $R_0$). This is highly unlikely in our experiment because we used the 4 probes technique to minimize contact resistance. Therefore, the almost flat shape of the curve $R/R_0$ Vs strain observed in Figure 3c should be attributed to the occurrence of mechanical sintering during stretching. The mechanical sintering phenomenon can be attributed to the rupture of the oxide shell and the reflow of liquid metal upon stretching, resulting in new percolation pathways that leads to a decreased value of resistivity.[37,40] At the same time, during stretching, freshly-formed surface area gets exposed to air and oxidizes, increasing the total resistivity of the EGaIn. This behavior explains that upon unloading the strain, the sample initiated reoxidation, causing the resistance to be higher compared to the stretching phase. This mechanism explains the hysteresis loop observed in Figure 3c, in which the line displays a slightly higher resistance during the contraction half-part of the cycle compared to the stretching one. As the number of stretching cycles increases, the accumulation of the oxide shell in the system becomes more pronounced and the overall resistance of the line increases. This behavior could be easily avoided by encapsulating the lines or by measuring them in inert atmosphere, and takes nothing aways from the claim of excellent stretchability of our plated EGaIn lines.

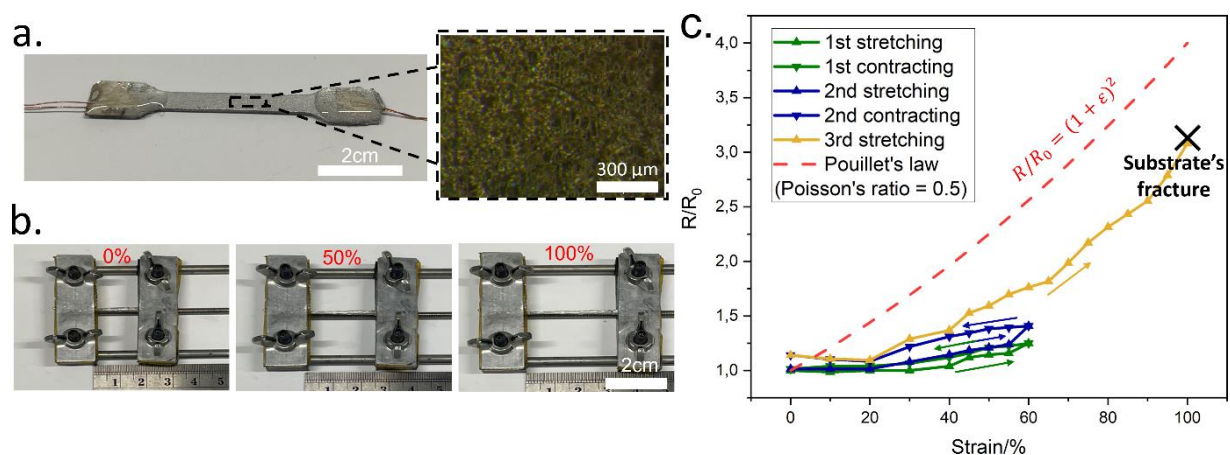

Figure 3: a) Photograph of an EGaIn line electroplated on top of a dogbone-shaped PDMS substrate, and close-up view of its surface morphology obtained under an optical microscope. b) Photographs of a deposited EGaIn sample on the PDMS substrate, clamped into a stretching device and subjected to 0%, 50% and 100% strain. c) Relative resistance variation ($R/R_0$) of the deposited EGaIn subjected to iterative reversible stretching up to fracture at 100% strain. The red dash line corresponds to Pouillet's law (assuming a Possion's ratio of PDMS = 0.5).

Due to the increasing oxidation of the un-encapsulated lines to assess the fatigue behavior of the deposited EGaIn lines, a cycling test was conducted under inert atmosphere. An EGaIn sample was placed inside an Ar-filled glove box and subjected to repeated strain of 75%. The obtained results indicate that the resistance remained relatively constant, with less than 20% increase after 300 cycles. In contrast, when the same experiment was realized in ambient conditions, a 268% increase in resistance was obtained. This experiment suggests that our EGaIn material exhibits excellent stretchability and demonstrates a high tolerance to fatigue.

*Electrodeposition of EGaIn with sub-micrometer resolution*

The "bottom-up" approach of the electrodeposition process removes the miniaturization constraints imposed by the high surface tension of liquid metal. Moreover, the patterning complexity or our process lies on the seed layer, which offers a higher degree of flexibility in the selection of the patterning method. To prove this claim, we fabricated stretchable PDMS substrates with 300 nm nanometer-wide trenches that could be filled with EGaIn. Fabrication of the substrates was done using a simple and low-cost method based on nanoimprinting and soft lithography-inspired "lift off" (shown in Figure 4a). Liquid PDMS was first carefully poured and cured on a patterned Si nanoimprinting mold functionalized with an anti-adhesion coating (1). After solidification, the PDMS replica was peeled off the Si wafer in the direction of the trenches (2) and laminated on a glass slide to act as stretchable substrate (3). Next, the trenched PDMS film was sputtered with a thin layer of gold (4). The sputtered gold on top of the trenches was subsequently removed employing a "lift off" step inspired in soft lithography (5). Adhesive tape was attached on top of the PDMS substrate, and after dwelling for 12 hours to improve adhesion, carefully peeled off in the direction of the trenches (5). This yielded trenches of PDMS with gold seed layers at their bottom. A droplet of silver paste was added to one end of the substrate, which seeped into the trenches and made electrical contact with the gold seed layers. Subsequently, electrodeposition of In and Ga was carried out on the sub-micron trenches using the acetonitrile-based electrolyte leading to EGaIn (500 mM of $GaCl_3$, 87.5 mM of $InCl_3$). The application of a particular voltage sequence was required to successfully fill the small trenches: a pulsed repeating sequence of −3.00 V vs. $Ag^+$/Ag for 10 ms followed by −2.00 V vs. $Ag^+$/Ag for 5 s for a total duration of 6 minutes at room temperature. Interestingly, for very short depositions (30 s), individual spheres were observed at discrete locations inside the trenches (SEM images are shown in Figure S2). It was postulated that the higher voltage (−3.00 V) pulse ensures nucleation over the entire seed layers inside the trenches, whereas the longer "normal" voltage (−2.00 V) induces growth of the nuclei. The morphology of the attained final sample is shown in Figure 4c. The SEM images reveal that the trenches are indeed filled up with EGaIn with the right composition (Figure 4d), and a smooth morphology rather than as a collection of spheres. This might be due to the mechanical sintering effect produced by swelling and shrinking of the PDMS trenches during and after exposure to the electrolyte, and by handling the sample after deposition. The elemental maps of gallium (from EGaIn) and silicon (from the PDMS substrate) obtained by EDX show clear alternating lines. Indium produces fainter lines than gallium owing to its overall lower concentration. It is worth noting that the minimum dimension of the fabricated lines (i.e. 300 nm) was imposed by the Si mold that we had at hand in our laboratory but there is in principle no is no reason why the patterning resolution cannot go below 300 nm.

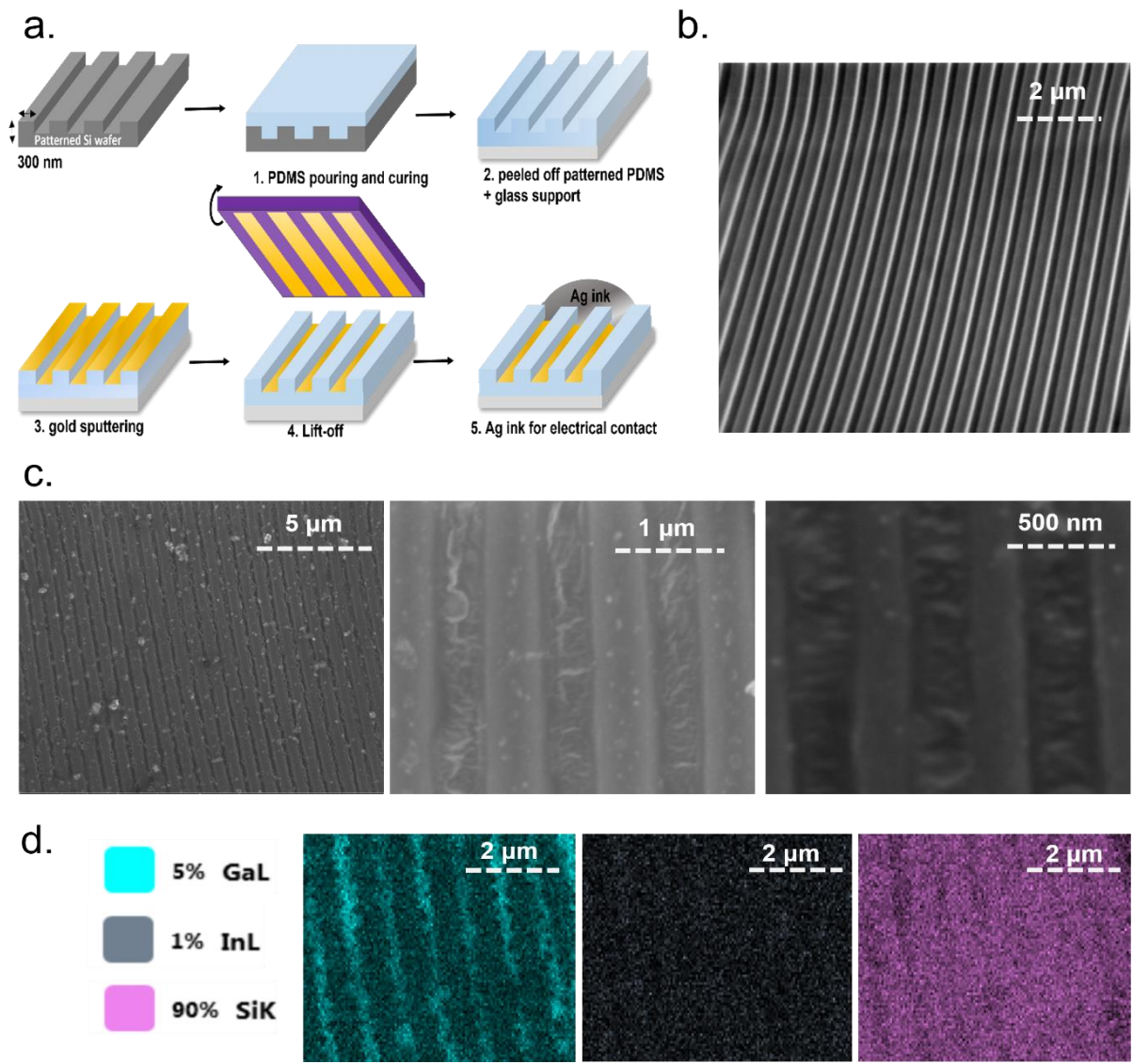

Figure 4: a) Sketch of stepwise fabrication of the sub-micron trenched elastomer substrates. b) SEM image of the resulting PDMS trenched substrate before the lift-off step (step 4 in panel a). c) SEM images at different magnifications of the same PDMS trenched substrate filled with EGaIn, electrodeposited by applying a pulsed repeating sequence of −3.00 V vs. Ag$^+$/Ag for 10 ms followed by −2.00 V vs. Ag$^+$/Ag for 5 s for a total duration of 6 minutes using an electrolyte composed acetonitrile, 500 mM of GaCl$_3$ and 87.5 mM of InCl$_3$. d) EDX elemental mapping of the sub-micron PDMS trenches filled with EGaIn.

*Filling high-aspect ratio through-elastomer vias by electrodeposition of EGaIn*

3D integration permits achieving electronics with enhanced integration density that leads to boosted functionality and computing power per device surface area. To connect different metallization layers in current Si integrated circuits technology, through-silicon vias (TSVs) filled with electroplated copper (termed the damascene process) are typically used. Electroplating (or electrodeposition) is an ideal technique to fill vias with high aspect ratios in a defect-less manner due to its bottom-up filling approach. Despite the high interest to extend the concept of 3D integration to soft electronics,[41–44] no real soft-electronics analogue of the damascene process exists. We demonstrate that our process for electrodeposition of EGaIn allows filling through-elastomer vias (TEVs) with high a high aspect ratio of 8:1 (to the best of our knowledge the highest aspect ratio reported so far for liquid metal vias is 5.7) and a diameter smaller than 500 μm.[44] TEVs permit connecting two stretchable EGaIn metallization layers to form an omnidirectional soft/stretchable electrical circuit. The stepwise process to fabricate the omnidirectional soft device is shown in Figure 5a and detailed in the Experimental section. Two parallel EGaIn lines were patterned on a PDMS substrate to conform the first metallization layer. The vias were created through a 4 mm-thick PDMS film (Figure S7), and were subsequently filled with EGaIn by electrodeposition (Figure 5b, c). After tuning the process (Figure S8), we could achieve densely filled vias ended in a mushroom cap shape (Figure 5b, c and Figure S9). Lastly, a surface mount LED was placed on the 4mm-thick PDMS film and connected with another EGaIn line to form the second metallization layer. The second metallization layer was finally encapsulated with another PDMS film to form a 3D circuit that was robust and mechanically compliant to different strain modes such as bending inward/outward around different axes, stretching, compressing and twisting (Figure 5d, movie S1).

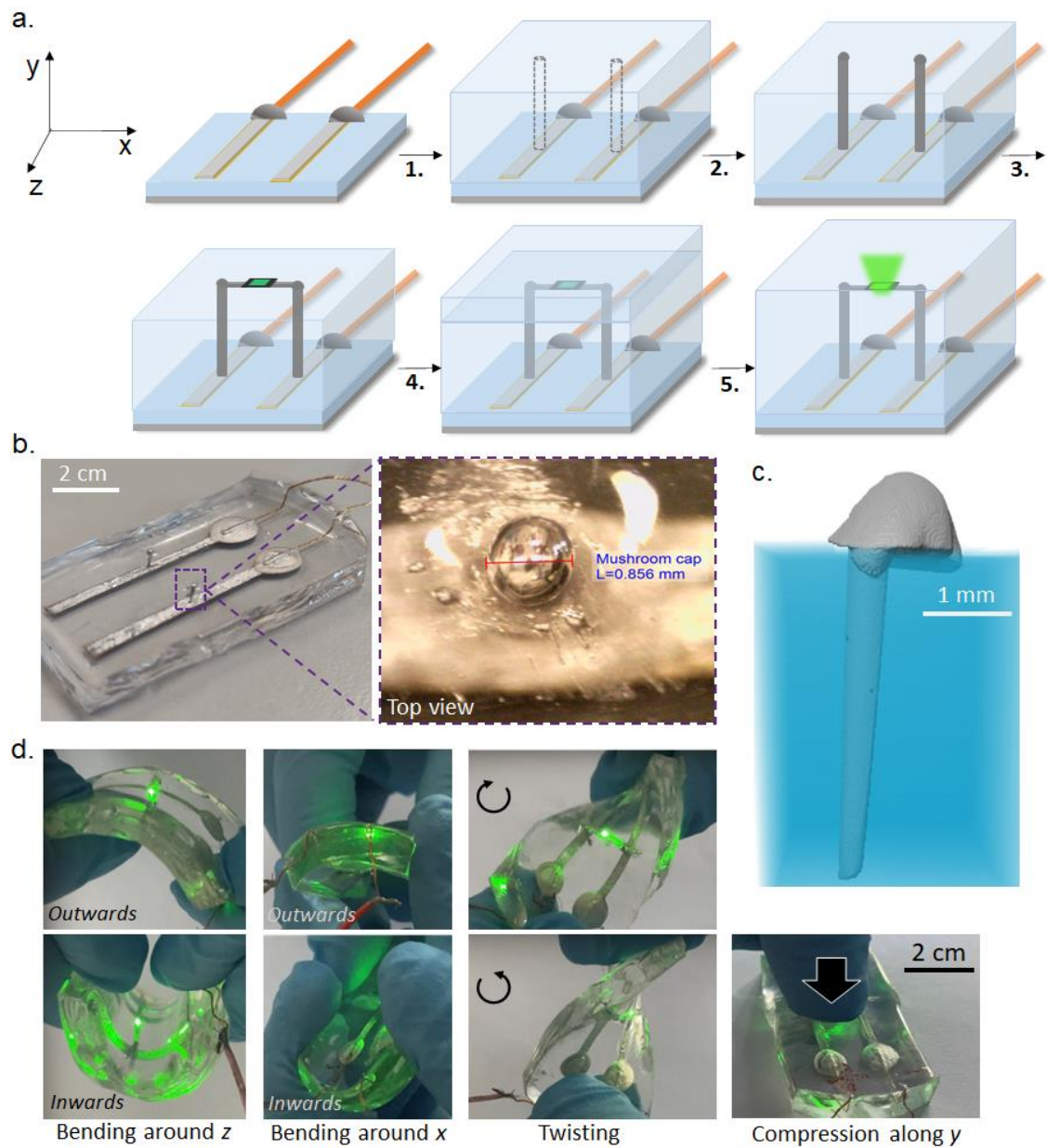

Figure 5: a) Sketch of the stepwise fabrication process of an omnidirectional soft/stretchable electrical circuit enabled by through-elastomer vias (TEVs) filled with EGaIn. The device consists of two parallel stretchable lines (1$^{st}$ metallization layer), two filled TEVs, a mounted LED light connected by a EGaIn line (2$^{nd}$ metallization layer), and a final encapsulation layer. b) Photograph of the device after filling the TEVs with a zoom-in top-view microscope picture of the liquid metal mashroom cap formed after electrodeposition. c) 3D visualization of X-ray Computed Tomography scans of a TEV densely filled with EGaIn by electrodeposition. d)

Photographs of an omnidirectional soft/stretchable electrical circuit showing how it remains operational (the LED remains on) when subjected to different representative strain modes: bending inward/outward around two different in-plane axes ($z$ and $x$ according to panel (a)), twisting clockwise and counterclockwise, and compression along the through-plane direction ($y$).

**Conclusions**

In summary, we report a new scalable "bottom-up" patterning technique for highly-integrated stretchable electronics based on the electrodeposition of EGaIn from a non-aqueous electrolyte composed of acetonitrile, $GaCl_3$ and $InCl_3$. We demonstrate that consistent electrodeposition of EGaIn can be achieved with a very high deposition efficiency (98.6 %). SEM-EDX analysis on lines deposited on a gold pre-patterned PDMS substrate indicated that electrodeposition of EGaIn leads to the formation of spheres, consisting of liquid metal cores with a thin gallium oxide shell. The stretchability of the substrate is leveraged to perform mechanical sintering that leads to the rupture of the shells and hence to electrically conducting lines with a smooth surface. The lines exhibited a high conductivity of $2.5 \times 10^6$ S m$^{-1}$ and could be stretched up to > 100% strain without significant decrease in conductivity. Furthermore, the conductivity remained stable even after 300 cycles at 75% strain. Sub-micrometer (300 nm) EGaIn regular structures were fabricated on PDMS, establishing a new milestone in the miniaturization of stretchable conductors. The capability of our method for the 3D integration of stretchable circuits is also demonstrated by filling high aspect ratio (8:1) vias, a process that could be considered as the soft matter analogue to the damascene process typically used in micro/nanoelectronics. Such integration capability is exemplified by a multilayer soft circuit in which a microLED integrated on a top layer is powered by tracks deposited on a bottom layer.

The 3D circuit remains operational while subjected to different strain modes (bending, stretching, compressing and twisting). Due to its scalability, simplicity and "bottom-up" nature, the presented proof-of-principle electrodeposition of EGaIn shows potential for the development of nanoscale 3D integration of conductor patterns for complex stretchable electronics devices.

**Experimental**

*Products*

Gallium(III) chloride ($GaCl_3$ anhydrous beads, 99.99%), indium(III) chloride ($InCl_3$, 99.99 %), silver nitrate ($AgNO_3$, 99.8 %), acetonitrile (anhydrous 99.98%), tetrabutylammonium chloride ([TBA][Cl], 99.5%) and tetrabutylammonium nitrate ([TBA][$NO_3$], 98.0%) were purchased from *Sigma-Aldrich* (Overijse, Belgium). Hydrochloric acid (*Analar Normapur*, 37%) was purchased from *VWR* (Leuven, Belgium). Ethanol and acetone (99 + %) were purchased from *Chem-Lab* (Zedelgem, Belgium). Silver conductive ink was purchased from *Fisher-Scientific* (Merelbeke, Belgium). Polydimethylsiloxane (PDMS, *Sylgard 184*) was purchased from *Dow Corning* (MI, USA). Tetrabutylammonium chloride was dried on a Schlenk line at 70 °C for 48 hours before use. All other chemicals, except for the silver ink and PDMS, were stored in an argon-filled glovebox, and used as received, without any further purification.

*Methods and devices*

*Electrodeposition*: For all electrochemical processes, a three-electrode setup was used. The system was controlled by an *Autolab PGSTAT 302N* potentiostat connected to a computer with *NOVA2* software. Electrochemical processes and measurements were consistently performed in an argon-filled glovebox with oxygen and moisture concentrations around 1 ppm. The working electrode (WE) consisted of various Au-coated (sputtered using an *Edwards S150* sputter coater) PDMS substrates. The counter electrode (CE) consisted of a glassy carbon plate whose area was at least five times larger than that of the WE. The reference electrode (RE) was a silver(I)/silver electrode, comprised of a glass tube filled with a solution of silver nitrate (10 mM) and tetraethylammonium nitrate (200 mM) in acetonitrile in which a platinum wire with a diameter of 1 mm was immersed. Before performing measurements, WEs and CEs were washed with ethanol and air-dried. For the rotating ring disk (RRDE) measurement, a gold disk/platinum ring tip was used (disk diameter = 5.5 mm, ring inner diameter = 8.5 mm, ring outer diameter = 6.5 mm) with a theoretical collection efficiency of 38.3 %. The tip was mounted in a pine research MSE rotator.

*PDMS substrate preparation for EGaIn stretchability test*: Polydimethylsiloxane (PDMS) was prepared by mixing the siloxane base oligomers and the curing agent with a mass ratio of 10:1. The mixed solution was poured into a dogbone mold and was cured at a constant temperature of 40 °C overnight. A plastic shadow mask was placed over the cured PDMS, and Au was sputtered for 90 seconds. The shadow mask was then removed, and EGaIn was electrodeposited on the exposed Au seed layer.

*Through PDMS vias preparation:* A layer of PDMS was cured on a glass substrate, on which two gold lines were sputtered. EGaIn was either manually added or electrodeposited on the gold lines. Next, an electrical contact was made between the liquid metal and copper wires using silver paste *Loctite ECI 1501 E&C*. After annealing the silver paste in an oven at 120 °C for 4 hours, a solid needle with a diameter of 500 μm was held vertically on top of each of the two patterned EGaIn lines, and then liquid (un-cured) PDMS was poured to encapsulate the lines and mold the shape of the needles. After the hardening of the PDMS layer, the needles were removed to create the vias. A picture of the sample with empty vias is shown in Figure S5. In the second step, the sample was immersed in the electrolyte. To fill the vias, electrical contact was made with the horizontal liquid metal wires (which served as the working electrode) using the two external copper wires. Electrodeposition of EGaIn was subsequently carried out by applying −2.00 V vs. $Ag^+/Ag$ for approximately 45 min. using an electrolyte composed of acetonitrile, 500 mM of $GaCl_3$ and 87.5 mM of $InCl_3$. Completion of the filling could be easily observed by the formation of small liquid metal mushrooms at the top of the vias (see Figure 5b, c and figure S9). Finally, a surface-mount LED was positioned between vias, connected by liquid metal lines and encapsulated with fresh PDMS.

*Characterization*

The morphology of the prepared deposits was studied by scanning electron microscopy (SEM). The SEM device used was a *FEI Nova 600 Nanolab* NanoSEM. For elemental analysis, energy-dispersive X-ray spectroscopy (EDX) (*Ametek EDAX*) was used. For all measurements, an acceleration voltage of 10 keV was applied. The resistance/resistivity measurements were performed using a 4-probe station *KeyFactor* Systems (HK) connected with a *Keithley 2000* multimeter. The thickness of the samples was determined using an optical microscope *Zeiss Axioskop 40 POL*. The relative resistance variation under strain was measured with a *Keithley 2000* multimeter. The sample was electrically contacted by bringing 4 Au-coated Cu wires in contact with the liquid metal line (Au helped the wettability of liquid metal on the wire for improved robustness), coating the wires with Ag paste (annealed for 60 min at 90 °C) and encapsulating the contact area with PDMS. Controlled strain was induced in a customized setup in which the sample was clamped in between two plates. The distance in between the plates was controlled by turning a screw, and measured with a ruler. The stretching cycles in inert atmosphere were done manually in an Ar-filled glovebox. The microstructure of vias and deposited lines were observed through optical microscopy and X-ray Computed Tomography (XCT). XCT was performed using a *TESCAN UniTOM XL* system (Czech Republic). The main scan parameters are summarized in Table 1.

Table 1. XCT utilized parameters.

| Scan parameters | |
|---|---|
| Parameter | Value |
| Acceleration voltage source (kV) | 50 |
| Target Power (W) | 15 |
| Filter | 1.5mm Al |
| Exposure time (ms) | 510 |
| Number of averages (#) | 3 |
| Projections per 360°(#) | 600 |
| Source-to-object distance (mm) | 30 |
| Source-to-detector distance (mm) | 700 |
| Magnification (X) | 23.3 |
| Voxel size (µm) | 6.43 |

**Acknowledgements**

WM thanks the Research Foundation Flanders (FWO) for supporting the research by the PhD grant (1SB8319N). BK: The research was further supported by the European Research Council (ERC) under the European Union's Horizon 2020 research and innovation programme: grant agreement 948922 – 3DALIGN. We acknowledge the FWO large infrastructure project (I013518N) and The KU Leuven XCT Core Facility for the 3D image acquisition and quantitative post-processing tools (www.xct.kuleuven.be). Lastly the authors thank Bogumila Kutrzeba Kotowska and Tibor Kuna for providing a patterned Si wafer.

**Conflict of interest**

There are no conflicts of interest to declare.

**Supporting Information**

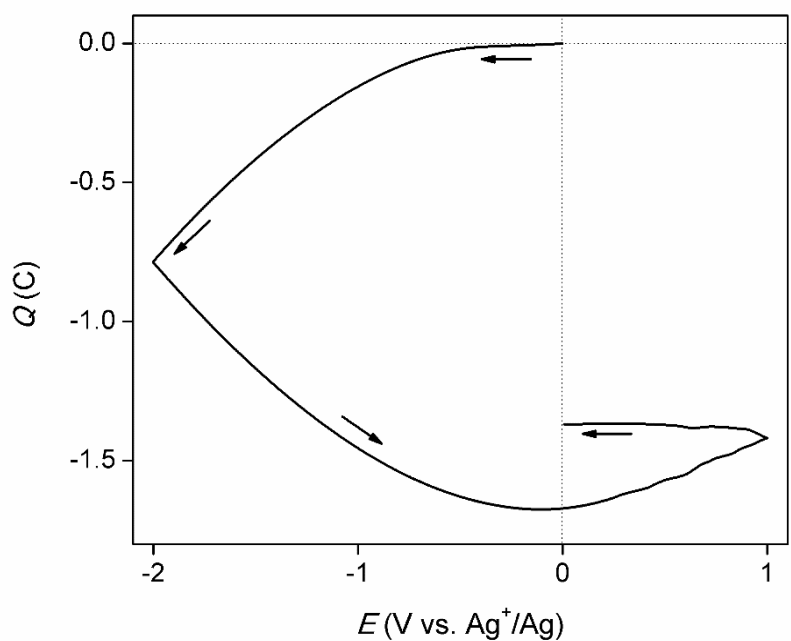

Figure S1: Charge vs. potential plot, derived from the CV shown in Figure 1c.

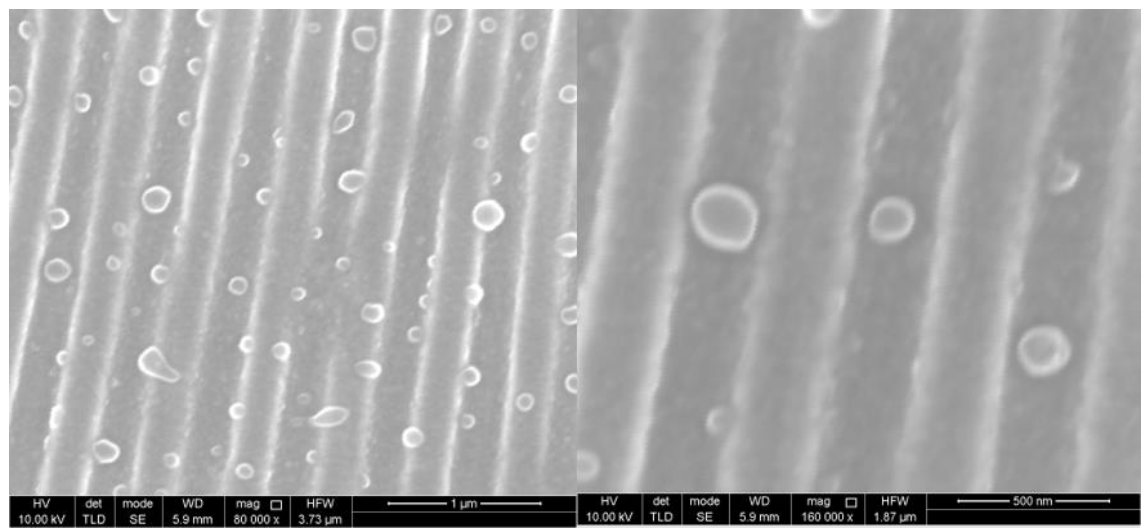

Figure S2: SEM images of sub-micron PDMS trenches filled with EGaIn spheres attained by electrodeposition −2.00 V vs. Ag$^+$/Ag for 30 s. using an electrolyte composed acetonitrile, 500 mM of GaCl$_3$, 62.5 mM of InCl$_3$.

Poisson's ratio for the conservation of volume of incompressible materials under strain:

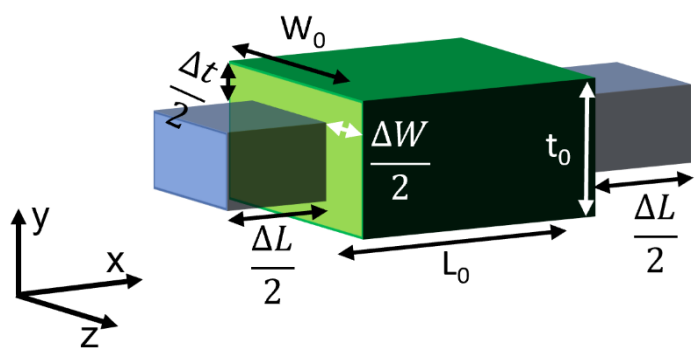

Figure S3: Sketch of the change of dimensions of a solid prisme subjected to uniaxial strain along its long axis (*x*). The dimensions along the perpendicular directions (*y* and *z*) decrease according to the Poissons's ratio ν of the material.

The definition of the Poisson's ratio ν of an isotropic material under an infinitesimally small deformation along the *x* axis, *dx,* is:

$$\nu = -\frac{\frac{dy}{y}}{\frac{dx}{x}} = -\frac{\frac{dz}{z}}{\frac{dx}{x}} \qquad (eq.\ S1)$$

Hence, the relative deformation along the perpendicular directions is given by $dy/y = dz/z = -\nu\, dx/x$. Integrating over the full extent of the deformation according to Figure S3 (example for *z*):

$$\int_{L_0}^{L_0+\Delta L} -\nu \frac{dx}{x} = \int_{W_0}^{W_0+\Delta W} \frac{dz}{z} \qquad (eq.\ S2)$$

Leads to the expression of the width of the strain solid:

$$W = W_0 + \Delta W = W_0 \left(1 + \frac{\Delta L}{L_0}\right)^{-\nu} \qquad (eq.\ S3)$$

An equivalent expression can be deduced for the *y* direction leading to the height of the strained solid: $t = t_0 + \Delta t = t_0 \left(1 + \frac{\Delta L}{L_0}\right)^{-\nu}$.

The volume of the solid under strain is $V = V_0 + \Delta V = (L_0 + \Delta L)(W_0 + \Delta W)(t_0 + \Delta t)$. Substituting eq. S3 and its equivalent for the *y* direction, and defining the strain as $\varepsilon = \Delta L/L_0$:

$$V = V_0 + \Delta V = L_0 W_0 t_0 (1 + \varepsilon)^{1-2\nu} = V_0 (1 + \varepsilon)^{1-2\nu} \qquad (eq.\ S4)$$

From the equation above, it can be inferred that for an incompressible material, for which the volume must remain constant under strain, *ΔV* must be equal to 0. This condition requires the power term (1 - 2 ν) = 0, and hence the Poisson's ratio ν = 0.5.

Deduction of the theoretical change of resistance under strain according to the Pouillet's law:

According to the Pouillet's law, the resistance $R_0$ of a wire is proportional to its resistivity $\rho$ and its length $L_0$, and inversely proportional to its cross-sectional area $S_0$ according to the following equation:

$$R_0 = \rho \frac{L_0}{S_0} \qquad (\text{eq. S5})$$

Upon stretching the wire along its length direction by an amount equal to $\Delta L$ (corresponding to a strain $\varepsilon = \Delta L/L_0$), the length would increase from $L_0$ to $L = L_0 + \Delta L$, and the cross-sectional area would decrease from $S_0$ to $S = S_0 + \Delta S$ (with $\Delta S < 0$) according to the Poisson's ratio of the material. Hence the resistance upon stretching becomes:

$$R = R_0 + \Delta R = \rho \frac{L_0 + \Delta L}{S_0 + \Delta S} \qquad (\text{eq. S6})$$

Expressing the cross-sectional surface area of the stretched wire in function of its stretched volume: $S_0 - \Delta S = (V_0 + \Delta V)/(L_0 + \Delta L)$. For an incompressible material $\Delta V = 0$, then $S_0 + \Delta S = V_0/(L_0 + \Delta L) = S_0 L_0/(L_0 + \Delta L)$ and:

$$R = R_0 + \Delta R = \rho \frac{(L_0 + \Delta L)^2}{S_0 L_0} = \rho \frac{L_0}{S_0}(1 + \Delta L/L_0)^2 = R_0(1 + \varepsilon)^2 \qquad (\text{eq. S7})$$

And the relative evolution of resistance under uniaxial stress becomes:

$$\frac{R}{R_0} = (1 + \varepsilon)^2 \qquad (\text{eq. S8})$$

Another deduction based on the Poisson's ratio can be found in the supplementary material of ref.[1]

Adding the effect of a constant contact resistance $R_c$ to the equation above is straight forward. $R_0$ becomes $R_0 = \rho \frac{L_0}{S_0} + R_c$, and eq. S8 can be rewritten as:

$$\frac{R}{R_0} = \frac{R_c}{R_0} + \left(1 - \frac{R_c}{R_0}\right)(1+\varepsilon)^2 \qquad \text{(eq. S9)}$$

According to Eq. S9, a non-negligible constant resistance would smooth out the increasing character of *R/R₀* as depicted in Figure S4.

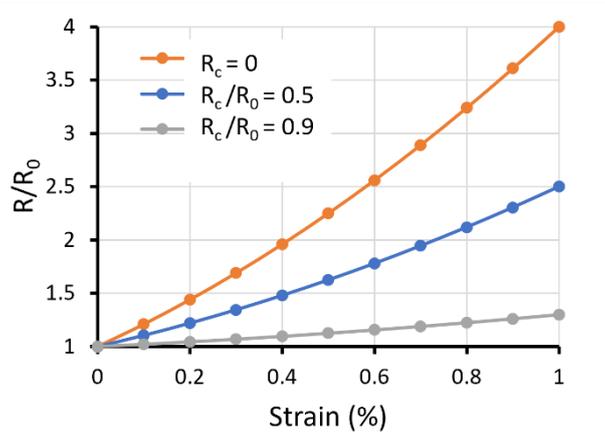

Figure S4: Value of the resistance of a wire under strain (relative to the unstrained resistance) according to the Poulliet's law for different relative values of contact resistance.

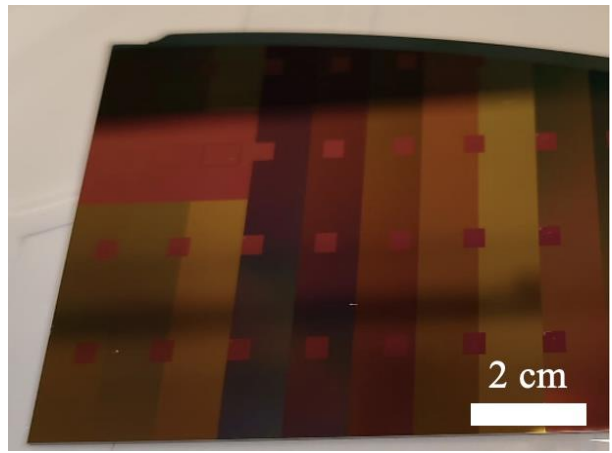

Figure S5: Picture of a nanopatterned Si mold for nanoimprint lithography.

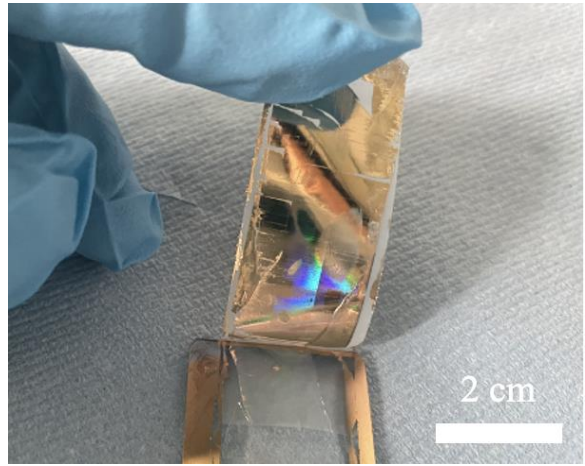

Figure S6: Picture of a piece of tape after removing the sputtered Au film from the top of the PDMS trenches (corresponding to Figure 4a, step #4). The diffracted colors of the Au film on the tape matched those of the Si mold used to pattern the PDMS (shown in Figure S5).

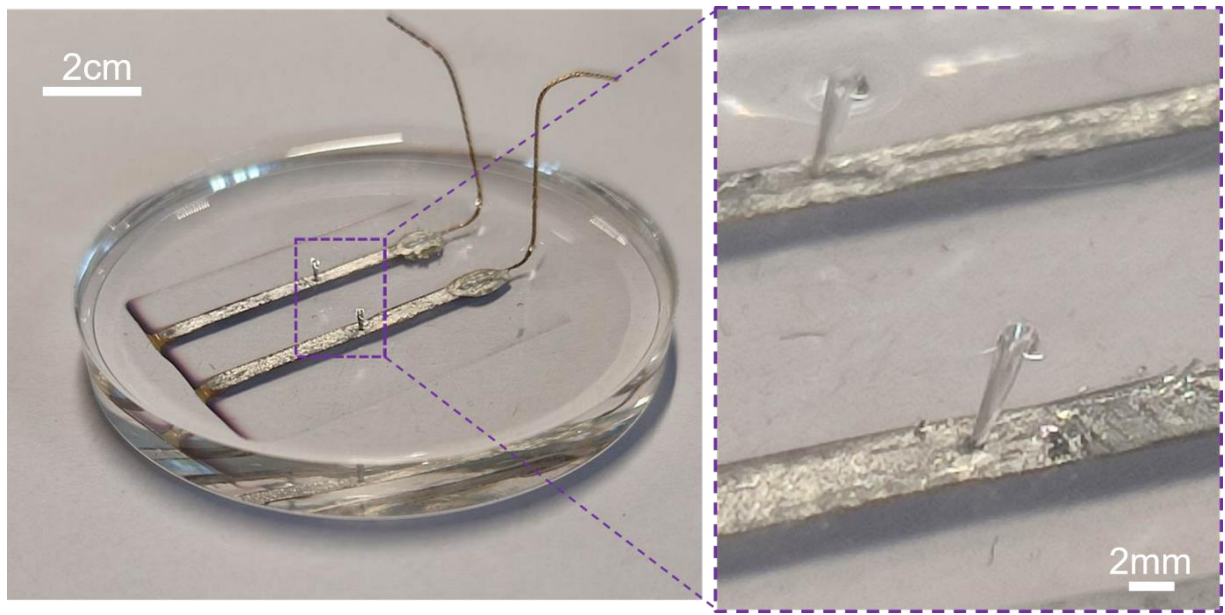

Figure S7: Pictures of an incomplete omnidirectional soft/stretchable electrical circuit, taken before filling the through-elastomer vias (corresponding to Figure 5a, step #2).

Figure S8 shows filled and partially filled through-elastomer vias (TEVs). All TEVs exhibited electrical connection, yet differences in filling quality were observed. This might be due to poor wetting of the acetonitrile-based electrolyte on PDMS that prevents the electrolyte to fill completely the vias (indeed high quality vias such as shown on the left were typically obtained by injection of electrolyte into the empty via, prior to immersion of the sample in the electrolyte bath); dewetting of the liquid metal from the PDMS walls within the vias; or irregularities during the electrodeposition process, such as the presence of small bubbles in the vias that prevent smooth filling.

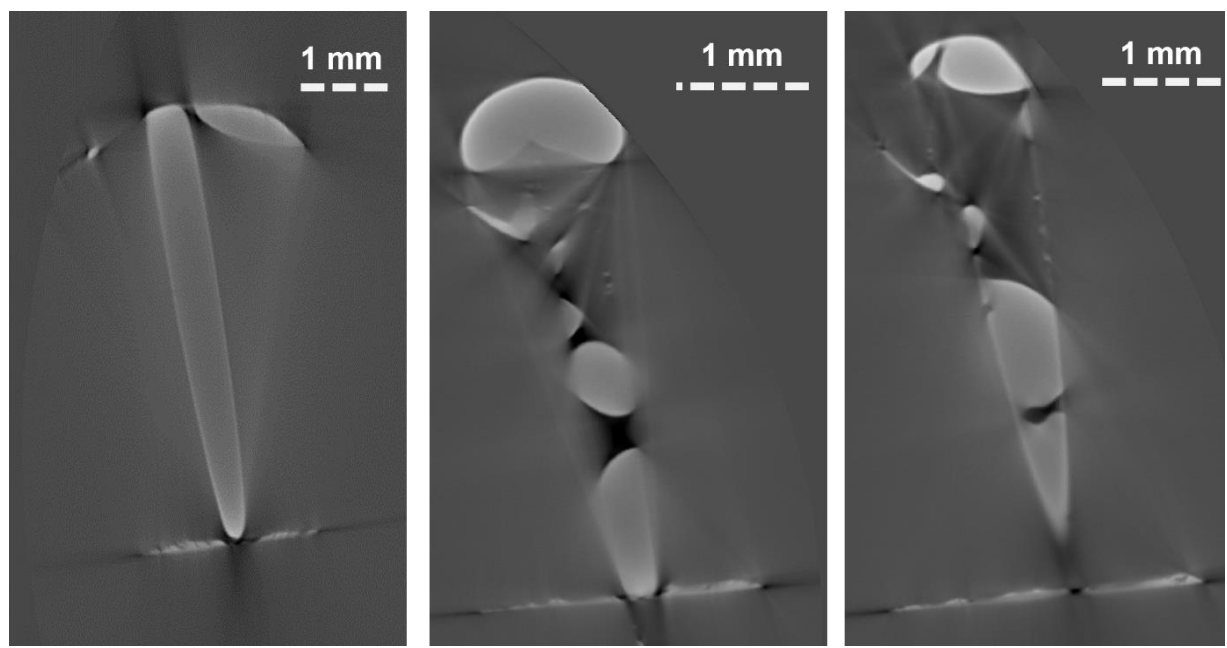

Figure S8: X-ray Computed Tomography scans of (partially) filled vias.

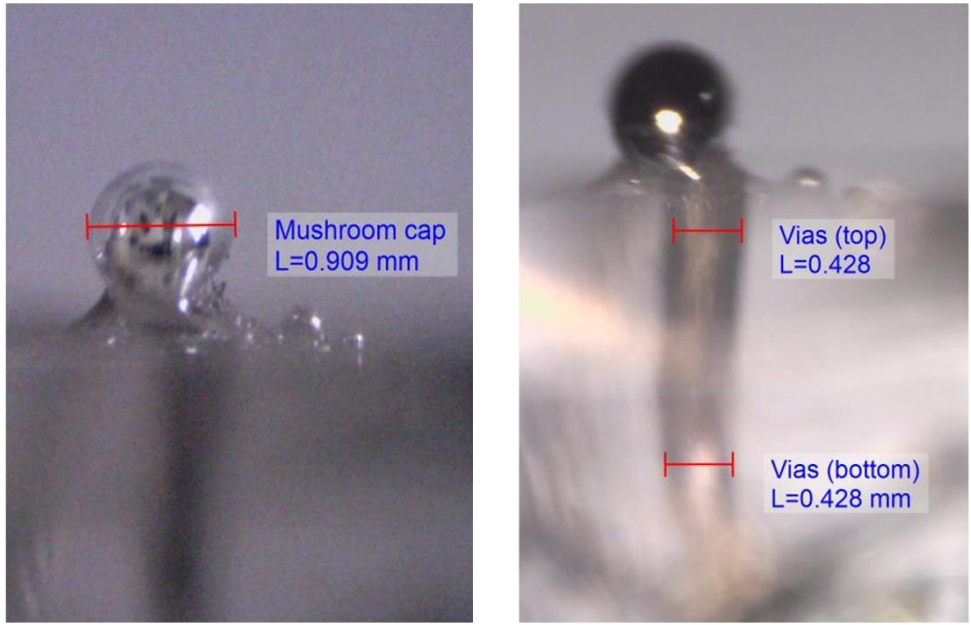

Figure S9: Optical microscope images of the cross section of a fully filled via with a mushroom cap.